\documentclass[onecolumn,showpacs,preprintnumbers,amsmath,amssymb,eqsecnum]{revtex4}
\usepackage{dcolumn}
\usepackage{bm}
\usepackage[dvips]{graphicx}
\usepackage{wrapfig}
\usepackage{psfrag}
\begin{document}

%Преобразуем нумерацию формул
%\renewcommand{\theequation}{\arabic{equation}}

\def\sh{\mathop{\rm sh}\nolimits}
\def\ch{\mathop{\rm ch}\nolimits}
\def\var{\mathop{\rm var}}\def\exp{\mathop{\rm exp}\nolimits}
\def\Re{\mathop{\rm Re}\nolimits}
\def\Sp{\mathop{\rm Sp}\nolimits}
\def\kp{\mathop{\text{\ae}}\nolimits}
\def\bk{{\bf {k}}}
\def\bp{{\bf {p}}}
\def\bq{{\bf {q}}}
\def\lra{\mathop{\longrightarrow}}
\def\Const{\mathop{\rm Const}\nolimits}
\def\sh{\mathop{\rm sh}\nolimits}
\def\ch{\mathop{\rm ch}\nolimits}
\def\var{\mathop{\rm var}}
\def\mK{\mathop{{\mathfrak {K}}}\nolimits}
\def\mR{\mathop{{\mathfrak {R}}}\nolimits}
\def\mv{\mathop{{\mathfrak {v}}}\nolimits}
\def\mV{\mathop{{\mathfrak {V}}}\nolimits}
\def\mD{\mathop{{\mathfrak {D}}}\nolimits}
\def\mN{\mathop{{\mathfrak {N}}}\nolimits}
\def\mS{\mathop{{\mathfrak {S}}}\nolimits}

\newcommand\ve[1]{{\mathbf{#1}}}

\def\Re{\mbox {Re}}
\newcommand{\Z}{\mathbb{Z}}
\newcommand{\R}{\mathbb{R}}
\def\mK{\mathop{{\mathfrak {K}}}\nolimits}
\def\mk{\mathop{{\mathfrak {k}}}\nolimits}
\def\mR{\mathop{{\mathfrak {R}}}\nolimits}
\def\mv{\mathop{{\mathfrak {v}}}\nolimits}
\def\mV{\mathop{{\mathfrak {V}}}\nolimits}
\def\mD{\mathop{{\mathfrak {D}}}\nolimits}
\def\mN{\mathop{{\mathfrak {N}}}\nolimits}
\def\ml{\mathop{{\mathfrak {l}}}\nolimits}
\def\mf{\mathop{{\mathfrak {f}}}\nolimits}
\newcommand{\ccm}{{\cal M}}
\newcommand{\cE}{{\cal E}}
\newcommand{\cV}{{\cal V}}
\newcommand{\cI}{{\cal I}}
\newcommand{\cR}{{\cal R}}
\newcommand{\cK}{{\cal K}}
\newcommand{\cH}{{\cal H}}
\newcommand{\cW}{{\cal W}}

\def\br{\mathop{{\bf {r}}}\nolimits}
\def\bS{\mathop{{\bf {S}}}\nolimits}
\def\bA{\mathop{{\bf {A}}}\nolimits}
\def\bJ{\mathop{{\bf {J}}}\nolimits}
\def\bn{\mathop{{\bf {n}}}\nolimits}
\def\bg{\mathop{{\bf {g}}}\nolimits}
\def\bv{\mathop{{\bf {v}}}\nolimits}
\def\be{\mathop{{\bf {e}}}\nolimits}
\def\bp{\mathop{{\bf {p}}}\nolimits}
\def\bz{\mathop{{\bf {z}}}\nolimits}
\def\bbf{\mathop{{\bf {f}}}\nolimits}
\def\bb{\mathop{{\bf {b}}}\nolimits}
\def\ba{\mathop{{\bf {a}}}\nolimits}
\def\bx{\mathop{{\bf {x}}}\nolimits}
\def\by{\mathop{{\bf {y}}}\nolimits}
\def\br{\mathop{{\bf {r}}}\nolimits}
\def\bs{\mathop{{\bf {s}}}\nolimits}
\def\bH{\mathop{{\bf {H}}}\nolimits}
\def\bk{\mathop{{\bf {k}}}\nolimits}
\def\be{\mathop{{\bf {e}}}\nolimits}
\def\bnul{\mathop{{\bf {0}}}\nolimits}
\def\bq{{\bf {q}}}

\newcommand{\oV}{\overline{V}}
\newcommand{\vkp}{\varkappa}
\newcommand{\os}{\overline{s}}
\newcommand{\opsi}{\overline{\psi}}
\newcommand{\ov}{\overline{v}}
\newcommand{\oW}{\overline{W}}
\newcommand{\oPhi}{\overline{\Phi}}

\def\mI{\mathop{{\mathfrak {I}}}\nolimits}
\def\mA{\mathop{{\mathfrak {A}}}\nolimits}

\def\st{\mathop{\rm st}\nolimits}
\def\tr{\mathop{\rm tr}\nolimits}
\def\sign{\mathop{\rm sign}\nolimits}
\def\d{\mathop{\rm d}\nolimits}
\def\const{\mathop{\rm const}\nolimits}
\def\O{\mathop{\rm O}\nolimits}
\def\Spin{\mathop{\rm Spin}\nolimits}
\def\exp{\mathop{\rm exp}\nolimits}
\def\SU{\mathop{\rm SU}\nolimits}
\def\mU{\mathop{{\mathfrak {U}}}\nolimits}
\newcommand{\cU}{{\cal U}}
\newcommand{\cD}{{\cal D}}

\def\mI{\mathop{{\mathfrak {I}}}\nolimits}
\def\mA{\mathop{{\mathfrak {A}}}\nolimits}
\def\mU{\mathop{{\mathfrak {U}}}\nolimits}

\def\st{\mathop{\rm st}\nolimits}
\def\tr{\mathop{\rm tr}\nolimits}
\def\sign{\mathop{\rm sign}\nolimits}
\def\d{\mathop{\rm d}\nolimits}
\def\const{\mathop{\rm const}\nolimits}
\def\O{\mathop{\rm O}\nolimits}
\def\Spin{\mathop{\rm Spin}\nolimits}
\def\exp{\mathop{\rm exp}\nolimits}

\title{Existence of an effective fermion vertex to lattice gravity}

\author {S.N. Vergeles\vspace*{4mm}\footnote{{e-mail:vergeles@itp.ac.ru}}}

\affiliation{Landau Institute for Theoretical Physics,
Russian Academy of Sciences,
Chernogolovka, Moscow region, 142432 Russia \linebreak
and   \linebreak
Moscow Institute of Physics and Technology, Department
of Theoretical Physics, Dolgoprudnyj, Moskow region,
141707 Russia}

\begin{abstract} It is shown that an effective fermion vertex arises to lattice gravity coupled with
fermions. The vertices are associated with gravitational instantons, much as the effective fermion vertices arising 
due to the existence of fermion zero modes associated with instantons
in the Yang-Mills theory.
\end{abstract}

\pacs{11.15.-q, 11.15.Ha}

\maketitle

\section{Introduction}

It has been shown many years ago that an effective chiral fermion vertex
associated with each instanton in the Yang-Mills theory coupled with massless Dirac fermions arises
\cite{1}-\cite{5}. The physics is as follows. Consider 4D Yang-Mills theory with the gauge group $SU(2)$ coupled with N massless Dirac fields $\Psi_i$, $i=1,\ldots,N$. Let's express the partition function for  Euclidean signature as the path integral over Yang-Mills and Dirac fields. The pure classical Yang-Mills equations have nontrivial topological solutions with finite action --- instantons \cite{6}. The Dirac massless operator  possess fermion zero modes, one for each Dirac field  (if the instanton topological charge is $q=\pm1$) which are localized  in the vicinity of instanton \cite{7}, \cite{1}, \cite{2}. The Dirac zero modes have definite chirality depending on the instanton topological charge $q=\pm1$.
The zero modes, in turn, create an effective chiral fermion vertex.
For definiteness, let's presume that the instanton is located in the vicinity of a point $x$ and its topological 
charge is $q=1$. Then the effective vertex (for Minkowski signature) is written as
\begin{gather}
V=\varkappa e^{-8\pi^2/g^2}\det_{st}\big\{\overline{\Psi}_{sR}(x)\Psi_{tL}(x)\big\}.
\label{10}
\end{gather}
Here the indices $s,t=1,\ldots,N$  enumerate the "flavor", the quantity $8\pi^2/g^2$ is the instanton action.
The arguments leading to the result (\ref{10}) are as follows. Let's consider fermion functional integral in
the external gauge field for vacuum-to-vacuum transition amplitude (with zeroth fermion sources).
The integral is proportional to $\{\det (i\hat{\cal{D}})\}$, where $(i\hat{\cal{D}})$ is the Dirac operator 
in the external gauge field. Since in the considered case the Dirac operator has zero eigenvalues, so the amplitude
is equal to zero. Now let's introduce the sources in the integral, so that the Dirac part of action
$S_{\Psi}=\int\d^{(4)}x\,\Psi^{\dag}i\hat{\cal{D}}\Psi$ becomes equal to 
$S_{\Psi}=\int\d^{(4)}x\,\Psi^{\dag}\{i\hat{\cal{D}}+J\}\Psi$. In this case the fermionic functional integral
gives us
\begin{gather}
\det\{i\hat{\cal{D}}+J\},
\label{20}
\end{gather}
which is not equal to zero since the lowest eigenvalues will become different from zero and they
are  of the order of $\sim J$.
Evidently, the factor of the determinant (\ref{20}) describing the lowest eigenvalues is equal to
\begin{gather}
\sim\det_{st}j_{st}(x), 
\nonumber \\
 j_{st}(x)=\int\d^{(4)}y\Psi_0^{(x)\dag}(y)J_{st}(y)\Psi_0^{(x)}(y).
\label{30}
\end{gather}
The designation $\Psi_0^{(x)}(y)$ is used for zero mode associated with instanton located 
in the vicinity of a point $x$.  It can readily be understood that the fermionic functional integral (\ref{20}), (\ref{30}) has the same effect (up to a c-number factor) as the functional integral
\begin{gather}
\int D\Psi^{\dag}D\Psi\exp\left\{\int\d^{(4)}y\,\Psi^{\dag}\left(i\hat{\cal{D}}+J\right)\Psi\right\}\det_{st}\big\{\Psi_s^{\dag}(x)(1+\gamma^5)\Psi_t(x)\big\}.
\label{40}
\end{gather}
If this is the case, creation of the effective fermion vertex (\ref{10}) is established.  First of all the zero eigenmodes for all flavors conveniently reproduce the fermionic propagators connecting
$x$ with the sources $J(y)$, if $|y-x|\gg \rho$, where $\rho$ is the scale of instanton, because of $\Psi_0^{(x)}(y)\sim
|y-x|^{-3}$ for $|y-x|\gg \rho$.
Secondly, the fact that (\ref{30}) does have the same quantum selection properties as (\ref{10}) is also evident if we regularize
correctly fermionic functional integral. For example, one can take the regularized fermion fields as
\begin{gather}
\Psi(x)=\sum_{|\epsilon_N|<\Lambda}\eta_N\Psi_N(x), \quad i\hat{\cal{D}}\Psi_N(x)=\epsilon_N\Psi_N(x),
\label{50}
\end{gather}
where $\{\eta_N\}$ are new Grassmann variables and $\Lambda\longrightarrow\infty$. The fermion zero modes are present at this decomposition of fermi-fields.  The vertex
(\ref{10}) has nonzero matrix elements only between the states with
\begin{gather}
\Delta Q^5\equiv\int\d^{(4)}x\partial_{\mu}J^{5\mu}=\int\d^{(4)}x\,{\cal A}^5=2N.
\label{60}
\end{gather}
From now on ${\cal A}^5$ denotes axial gauge anomaly in Yang-Mills or gravity theory.
The quantity (\ref{30}) also describe the process (\ref{60}).

\section{A short description of lattice gravity and some necessary results}

It is necessary to sketch out the model of lattice gravity ( Euclidean signature) which is used here. A detailed description of the model is given in \cite{8}-\cite{10}. We use here the $\gamma^a$-matrices, $a=1,\ldots,4$, in spinor representation, so that $\gamma^5$ and $\sigma^{ab}=1/4[\gamma^a,\gamma^b]$ are diagonal and block-diagonal matrices,
correspondingly. The orientable 4-dimensional simplicial complex and its vertices are designated as $\mK$ and
$a_{\cV}$, the indices ${\cV}=1,2,\dots,\,{\mN}\rightarrow\infty$ and ${\cW}$ enumerate the vertices
and 4-simplices, correspondingly. We assume here that $\mK\approx{\boldsymbol{\R^4}}$ in a topological sense. It is necessary to use
the local enumeration of the vertices $a_{\cV}$ attached to a given
4-simplex: the all five vertices of a 4-simplex with index ${\cW}$
are enumerated as $a_{{\cW}i}$, $i=1,2,3,4,5$. The later notations with extra index  ${\cW}$
indicate that the corresponding quantities belong to the
4-simplex with index ${\cW}$. The Levi-Civita symbol with in pairs
different indexes $\varepsilon_{{\cW}ijklm}=\pm 1$ depending on
whether the order of vertices
$s^4_{\cW}=a_{{\cW}i}a_{{\cW}j}a_{{\cW}k}a_{{\cW}l}a_{{\cW}m}$ defines the
positive or negative orientation of 4-simplex $s^4_{\cW}$.
An element of the group $\Spin(4)$ and an element of the Clifford algebra
\begin{gather}
\Omega_{{\cW}ij}=\Omega^{-1}_{{\cW}ji}=\exp\left(\omega_{{\cW}ij}\right), \quad
\omega_{{\cW}ij}\equiv\frac{1}{2}\sigma^{ab}\omega^{ab}_{{\cW}ij},
\nonumber \\
\hat{e}_{{\cW}ij}=\hat{e}_{{\cW}ij}^{\dag}\equiv e^a_{{\cW}ij}\gamma^a\equiv-\Omega_{{\cW}ij}\hat{e}_{{\cW}ji}\Omega_{{\cW}ij}^{-1}.
\label{70}
\end{gather}
are assigned for each oriented 1-simplex $a_{{\cW}i}a_{{\cW}j}$.
The Dirac spinors $\Psi_{\cV}$ and $\Psi^{\dag}_{\cV}$ are assigned to each vertex $a_{\cV}$.
The used representation  realizes automatically the separation of a total gauge group into two sub-group:
$\Spin(4)\approx\Spin(4)_{(+)}\otimes\Spin(4)_{(-)}$. For example
\begin{gather}
\frac12\sigma^{ab}\omega^{ab}_{{\cW}ij}=
 \frac{i\sigma^{\alpha}}{2} \left(
\begin{array}{cc}
\omega^{\alpha}_{(+){\cW}ij} & 0  \\
0 & \omega^{\alpha}_{(-){\cW}ij} \\
\end{array} \right),
\nonumber \\
\omega^{\alpha}_{(\pm){\cW}ij}\equiv \left\{\pm\omega^{\alpha4}_{{\cW}ij}+\frac12\varepsilon_{\alpha\beta\gamma}
\omega^{\beta\gamma}_{{\cW}ij}\right\}.
\nonumber
\end{gather}
The considered lattice action has the form
\begin{gather}
\mA=\mA_g+\mA_{\Psi},
\nonumber \\
\mA_g=-\frac{2}{ 5\cdot
24\cdot l^2_P}\sum_{\cW}\sum_{i,j,k,l,m}\varepsilon_{{\cW}ijklm}\times
\nonumber \\
\times\tr\,\gamma^5
\Omega_{{\cW}mi}\Omega_{{\cW}ij}\Omega_{{\cW}jm}
\hat{e}_{{\cW}mk}\hat{e}_{{\cW}ml},
\nonumber \\
\mA_{\Psi}=-\frac{1}{5\cdot24^2}\sum_{\cW}\sum_{i,j,k,l,m}\varepsilon_{{\cW}ijklm}\times
\nonumber \\
\times\tr\,\gamma^5 \hat{\Theta}_{{\cW}mi}
\hat{e}_{{\cW}mj}\hat{e}_{{\cW}mk}\hat{e}_{{\cW}ml},
\nonumber \\
\hat{\Theta}_{{\cW}ij}=
\frac{i}{2}\gamma^a\left(\Psi^{\dag}_{{\cW}i}\gamma^a
\Omega_{{\cW}ij}\Psi_{{\cW}j}-\Psi^{\dag}_{{\cW}j}\Omega_{{\cW}ji}\gamma^a\Psi_{{\cW}i}\right)\equiv
\nonumber \\
\equiv\gamma^a\Theta^a_{{\cW}ij}.
\label{80}
\end{gather}
This action is invariant relative to the gauge transformations
\begin{gather}
\tilde{\Omega}_{{\cW}ij}=S_{{\cW}i}\Omega_{{\cW}ij}S^{-1}_{{\cW}j},
\quad
\tilde{e}_{{\cW}ij}=S_{{\cW}i}\,e_{{\cW}ij}\,S^{-1}_{{\cW}i},
\nonumber \\
\tilde{\Psi}_{{\cW}i}=S_{{\cW}i}\,\Psi_{A\,i},  \quad
\tilde{\Psi^{\dag}}_{{\cW}i}=\Psi^{\dag}_{{\cW}i}\,S^{-1}_{{\cW}i}
\nonumber \\
 S_{{\cW}i}\in\Spin(4).
\nonumber
\end{gather}

The action (\ref{80}) reduces to the continuum action of gravity in a four-dimensional Euclidean space in the limit of slowly varying fields. Consider a certain  $4D$ sub-complex of complex $\mK$ with the trivial topology of four-dimensional disk.
Realize geometrically this sub-complex in $\R^4$. Thus each vertex of the sub-complex  acquires
the coordinates $x^{\mu}$  which are the coordinates of the vertex image in $\R^4$:
\begin{gather}
x^{\mu}_{{\cW}i}=x^{\mu}_{\cV}\equiv x^{\mu}(a_{{\cW}i})\equiv x^{\mu}(a_{\cV}),
 \qquad \ \mu=1,\,2,\,3,\,4.
\nonumber
\end{gather}
 The four vectors
\begin{gather}
\d x^{\mu}_{{\cW}ji}\equiv x^{\mu}_{{\cW}i}-x^{\mu}_{{\cW}j},
 \quad i=1,\,2,\,3,\,4
\label{90}
\end{gather}
are linearly independent. In the continuous limit, the holonomy group elements (\ref{70}) are
close to the identity element, so that the quantities $\omega^{ab}_{ij}$
tend to zero being of the order of $O(\d x^{\mu})$.
Thus one can consider the following system of equation for $\omega_{{\cW}m\mu}$
\begin{gather}
\omega_{{\cW}m\mu}\,\d x^{\mu}_{{\cW}mi}=\omega_{{\cW}mi},  \quad
i=1,\,2,\,3,\,4\,.
\label{100}
\end{gather}
In this system of linear equation, the indices ${\cW}$ and $m$ are
fixed, the summation is carried out over the index $\mu$, and
index runs over all its values. Since the differentials (\ref{90}) are linearly independent, the quantities 
$\omega_{{\cW}m\mu}$
are defined uniquely. Thus equations (\ref{100}) define uniquely the connection 1-forms
$\omega^{ab}=\omega^{ab}_{\mu}\d x^{\mu}$.

If the fields $\omega_{\mu}$ smoothly depend on
the points belonging to the geometric realization of each
four-dimensional simplex, then the following formula is
valid up to $O\big((\d x)^2\big)$ inclusive
\begin{gather}
\Omega_{{\cW}mi}\,\Omega_{{\cW}ij}\,\Omega_{{\cW}jm}=
\exp\left[\frac{1}{2}\,\mR_{\mu\nu}(x_{{\cW}m})\d x^{\mu}_{{\cW}mi}\, \d
x^{\nu}_{{\cW}mj}\,\right]\,,
\nonumber
\end{gather}
where
\begin{gather}
\mR_{\mu\nu}=\partial_{\mu}\omega_{\nu}-\partial_{\nu}\omega_{\mu}+
[\omega_{\mu},\,\omega_{\nu}\,]\equiv
\frac12\sigma^{ab}\mR^{ab}_{\mu\nu}.
\nonumber
\end{gather}

In exact analogy with (\ref{100}), let us write out the following relations
for a tetrad field without explanations:
\begin{gather}
\hat{e}_{{\cW}m\mu}\,\d x^{\mu}_{{\cW}mi}=\hat{e}_{{\cW}mi}\longrightarrow
e^a=e^a_{\mu}\d x^{\mu}.
\nonumber
\end{gather}

Applying above-listed formulas  to the discrete action (\ref{80}) and
changing the summation to integration in the case smoothly varying field we obtain  the well known form of
gravity action:
\begin{gather}
\mA=\int\,\varepsilon_{abcd}\,\left\{-\frac{1}{l^2_P}\mR^{ab}\wedge e^c\wedge e^d-\frac{1}{6}\,\Theta^a\wedge e^b\ \wedge e^c\wedge e^d
\right\},
\nonumber \\
\gamma^a\Theta^a=\gamma^a\,\frac{i}{2}\,\left[
\Psi^{\dag}\gamma^a\,{\cal D}_{\mu}\,\Psi-
\left({\cal D}_{\mu}\,\Psi\right)^{\dag}\gamma^a\,\Psi
\right]\d x^{\mu},
\nonumber \\
{\cal D}_{\mu}\,\Psi=\partial_{\mu}
\Psi+\omega_{\mu}\Psi,
\quad
\Psi=
\left(
\begin{array}{c}
\phi \\
\chi \\
\end{array} \right).
\nonumber
\end{gather}

It has been proved  that  the
instanton-like self-dual solution does exist in the outlined lattice theory of pure gravity \cite{11}. In contrast to the well known Eguchi-Hanson solution to continuous Euclidean
Gravity \cite{12}, the lattice solution is asymptotically {\it{globally}} Euclidean, i.e., the 
boundary of the space as $r\longrightarrow\infty$ is $S^3=SU(2)$. This solution satisfies 
the general equations
\begin{gather}
\delta\mA_g/\delta\omega^{\alpha}_{(\pm){\cW}mi}=0, \quad
\delta\mA_g/\delta e^a_{{\cW}mi}=0,
\nonumber
\end{gather}
as well as the duality equations
\begin{gather}
\omega^{\alpha}_{(-){\cW}mi}=0 \longleftrightarrow \Omega_{(-){\cW}mi}=1.
\nonumber
\end{gather}
The boundary conditions are as follows:
\begin{gather}
\Omega_{(+){\cW}ij}=-1,  \quad s^4_{\cW}\in\mk,
\nonumber \\
e^a_{{\cV}_1{\cV}_2}=\phi^a_{{\cV}_2}-\phi^a_{{\cV}_1}, \quad a_{{\cV}_1}a_{{\cV}_2}\in\mk, \quad  a_{{\cV}_1}a_{{\cV}_2}\notin\partial\mk,
\nonumber
\end{gather}
where $\mk\subset\mK$ is a finite sub-complex  containing the centre of instanton with the
boundary $\partial\mk\approx S^3$. At infinity ($r\gg\rho$) we have
\begin{gather}
e^a_{\mu}\d x^{\mu}=
 \left(
\begin{array}{c}
\frac{r}{2}\left(\sin\psi\d\theta-\sin\theta\cos\psi\d\varphi\right)  \\
\frac{r}{2}\left(\cos\psi\d\theta+\sin\theta\sin\psi\d\varphi\right)  \\
-\frac{r}{2}g(r)\left(\cos\theta\d\varphi+\d\psi\right) \\
g(r)^{-1}\d r \\
\end{array} \right),
\quad
g(r)=\sqrt{1-\frac{\rho^4}{r^4}}, 
\nonumber \\
\frac{i\sigma^{\alpha}}{2}\omega^{\alpha}_{(+)}\longrightarrow U^{-1}\d U, \quad 
\qquad
U=\exp\left(-\frac{i\sigma^3}{2}\varphi\right)\exp\left(\frac{i\sigma^2}{2}\theta\right)
\exp\left(-\frac{i\sigma^3}{2}\psi\right).
\nonumber
\end{gather}
Here  $(\theta,\,\varphi,\,\psi)$ are the Euler angles varying in the ranges
$0\leq\theta\leq\pi$, $0\leq\varphi\leq 2\pi$, $0\leq\psi\leq 4\pi$, and $(\theta,\,\varphi,\,\psi,\,r)$ are the coordinates
of Euclidean space $\R^4$.

It has been proved also  that  the lattice fermion zero mode associated with lattice instanton and 
localized in the vicinity of the instanton does exist \cite{13}. To demonstrate the statement, let's
consider the "half" of fermion action (\ref{80}) which describes the dynamics of right fermion field $\phi$:
\begin{gather}
\mA_{(\phi)}=-\frac{1}{5\cdot6\cdot24}\sum_{\cW}\sum_{i,j,k,l,m}\varepsilon_{{\cW}ijklm}\varepsilon_{abcd}\,\Theta^a_{(\phi){\cW}mi}
e^b_{{\cW}mj}e^c_{{\cW}mk}e^d_{{\cW}ml},
\nonumber \\
\Theta^a_{(\phi){\cW}ij}=
\frac{1}{2}\left(\chi^{\dag}_{{\cW}i}\sigma^a
\Omega_{(+){\cW}ij}\phi_{{\cW}j}+\phi^{\dag}_{{\cW}j}\Omega_{(+){\cW}ji}(\sigma^a)^{\dag}\chi_{{\cW}i}\right),
\nonumber \\
\sigma^a\equiv\left(\sigma^1,\,\sigma^2,\,\sigma^3,\,i\right).
\label{110}
\end{gather}
It is convenient to write the continuous variant of the introduced fermion lattice action (\ref{110})
in the form
\begin{gather}
\mA_{(\phi)}=\int\d^{(4)}x\left(\det e^b_{\lambda}\right){}\bigg\{\frac{1}{2}e^{\mu}_{a}\big[
\chi^{\dag}\sigma^a{\cal D}_{(+)\mu}\phi+c.c.
\big]\bigg\}, 
\nonumber \\
{\cal D}_{(+)\mu}\equiv \partial_{\mu}+\frac{i}{2}\sigma^{\alpha}\omega^{\alpha}_{(+)\mu}.
\nonumber
\end{gather} 
One must solve lattice equations
\begin{gather}
\delta\mA_{\Psi(+)}/\delta\phi_{\cV}=0, \quad \delta\mA_{\Psi(+)}/\delta\chi^{\dag}_{\cV}=0,
\label{120}
\end{gather}
as well as their complex conjugate equations,  against the background of instanton field. It is proved in \cite{13} that two localized solutions of Eqs. (\ref{120}) (fermion zero modes)
do exist. Here only the  asymptotic behavior of the fermion zero modes is interesting.
For the instanton located at the zero point the solutions are as  follows:
\begin{gather}
\phi_0=\frac{\Const}{r^3}\exp\left(\frac{i\sigma^3}{2}\psi\right)
\left(
\begin{array}{c}
\sqrt{\tan\left(\theta/2\right)}  \\
\sqrt{\cot\left(\theta/2\right)}  \\
\end{array} \right),  \quad r\gg\rho,
\nonumber \\
\overline{\phi}_0=\left(i\sigma^2\phi_0\right)^*.
\label{130}
\end{gather}
Here the upper index ${}^*$ means complex conjugation. For both solutions
\begin{gather}
\chi^{\dag}_{0\,\cV}=0, \quad \chi_{0\,\cV}=0 \quad \mbox{for all  vertices} \quad  a_{\cV}.
\nonumber
\end{gather}

It is known that two spinors $\phi$ and $\overline{\phi}=\left(i\sigma^2\phi\right)^*$ transform identically
under the gauge transformations $\Spin(4)_{(+)}$ or $SU(2)_{(+)}$ \cite{14}. 
So  the decomposition of the field $\phi$ analogous to that in (\ref{50}) must be as follows: 
\begin{gather}
\phi_{s\cV}=\eta_{s\,0}\,\phi_{0\cV}+\eta_{s\,0}^{\dag}\,\overline{\phi}_{0\cV}+(\mbox{ all the rest}).
\label{140}
\end{gather}
We have 
\begin{gather}
\sum_{\cV}\overline{\phi}_{0\cV}^T(i\sigma^2)\phi_{0\cV}=1, 
\qquad
\overline{\phi}_{0\cV}^T(i\sigma^2)\overline{\phi}_{0\cV}=\phi_{0\cV}^T(i\sigma^2)\phi_{0\cV}=0.
\nonumber
\end{gather}
The upper index ${}^T$ means matrix transformation.

\section{The effective fermion vertex associated with instanton in the lattice gravity theory}

{\bf 3.} Now we reproduce the logic of the Point {\bf 1} for the case of gravity. It is assumed
that the right fields $\phi_s$, $s=1,\ldots,N$ are introduced. To avoid the fermion
determinant vanishing, the fermion action (\ref{110}) is modified by adding  the sources term
\begin{gather}
\Delta\mA_{\Psi(+)}=\sum_{s,t,\cV}\{J_{s,t,\cV}\phi^T_{s,\cV}(i\sigma^2)\phi_{t,\cV}+c.c\}=
\nonumber \\
=\sum_{s,t,\cV}\left(J_{s,t,\cV}+J_{s,t,\cV}^*\right)\left[\overline{\phi}_{0\cV}^T(i\sigma^2)\phi_{0\cV}\right]
\left(\eta^{\dag}_{s0}\eta_{t0}+\eta^{\dag}_{t0}\eta_{s0}\right)+(\mbox{ all the rest}).
\nonumber
\end{gather}
 Then the fermion functional integration 
 leads to the expression (\ref{30}), where now one must use
\begin{gather}
j_{st}(x)=\sum_{s,t,\cV}\left\{\left(J_{s,t,\cV}+J_{s,t,\cV}^*\right)
\overline{\phi}^{(x)T}_{0\,\cV}(i\sigma^2)\phi_{0\,\cV}^{(x)}\right\},
\nonumber 
\end{gather}
which, in turn, leads to an effective fermion Lorentz invariant vertex (compare with (\ref{10}))
\begin{gather}
V\sim \det_{st}\Big\{\phi^T_s(x)\sigma^2\phi_t(x)+\phi^{\dag}_s(x)\sigma^2\phi^{\dag\,T}_t(x)\Big\}.
\label{150}
\end{gather}
The designation $\phi_{0\,\cV}^{(x)}$ is used for instanton zero mode solution  located 
in the vicinity of a point $x$ (the instanton centre). 

The result of the work  is given by the last formula.

There is a significant difference between Yang-Mills and gravity theories concerning instanton physics. For simplicity let's consider the case $N=1$.
In Yang-Mills theory the integral in (\ref{60}) is  $\int\d^{(4)}x\,{\cal A}^5=2$ for instanton.
The axial anomaly is present also for gravity theory, but in this case $\int_{r>\rho}\d^{(4)}x\,{\cal A}^5=1/4$ 
even if we use the continuum calculations up to the region where lattice instanton has singularity. Such calculation
seems to be incorrect. It is not clear also how one can count essentially lattice contribution into axial anomaly?
But there is a  positive argument for the fact that the lattice gravity instanton  leads to  $\Delta Q^5=0$.
Indeed, in the continuous limit for Minkowski signature, the action (\ref{110})  transforms into the action
\begin{gather}
\mA_{\phi}=\int\d^{(4)}x\left(\det e^b_{\lambda}\right)e^{\mu}_a\left\{\phi^{\dag}\sigma^a{\cal D}_{(+)\mu}\phi\right\},
\nonumber \\
\sigma^a=\left(1,\,\sigma^1,\,\sigma^2,\,\sigma^3\right),
\nonumber
\end{gather}
describing the dynamics of right relativistic fermions. This action can not generate $\Delta Q^5\neq 0$.

\section{Conclusion}

 More recently, the idea that neutrino masses and so neutrino oscillations are generated by gravity interaction is discussed (see, for example, \cite{15}, \cite{16}).

Note that the vertex (\ref{150}) generates neutrino oscillations. Thus the result of the work
is resonant  in a sense with the works cited above. 
 
We  emphasise that the effect does exist only for lattice gravity. The reason is that the gravitational
instanton as well as fermion zero mode exist only for lattice case but not for totally  continuum theory.

\end{document}